\documentclass[pra,aps,twocolumn,superscriptaddress,dvipsnames,amssymb,9pt,floatfix]{revtex4}
\usepackage{graphicx}
\usepackage{bm}
\usepackage{amsmath,hyperref}
\usepackage{amssymb}
\usepackage{amsfonts}
\usepackage{fancyhdr}
\usepackage{slashed}
\usepackage{latexsym,bbm}
\usepackage{amsthm}
\usepackage{float}
\usepackage{amsthm}
\usepackage{algorithm,algpseudocode}
\usepackage{enumerate}
\usepackage{comment}

\newcommand{\ket}[1]{\left |{#1} \right \rangle}

\usepackage{color}
\definecolor{blue}{rgb}{0,0.2,1}

\definecolor{green}{rgb}{0,1,0}

\definecolor{red}{rgb}{0.9,0,0}

\newcommand{\Ord}[1]{\mathcal{O}\left( #1 \right)}

\theoremstyle{plain}

\newtheorem{theorem}{Theorem}

\newtheorem{lemma}{Lemma}

\newtheorem{defn}{Definition}

\def\be{\begin{eqnarray}}
\def\ee{\end{eqnarray}}

\usepackage{color}
\definecolor{Pr}{rgb}{0.4,0.3,0.9}

\begin{document}

\title{Learning Restricted Boltzmann Machines with greedy quantum search}
\author{Liming Zhao}\email{zlm@swjtu.edu.cn}
\affiliation{The School of Information Science and Technology, Southwest Jiaotong University, Chengdu, China 610031}

\author{Aman Agrawal}
\email{amanagr@iitk.ac.in}
\affiliation{Indian Institute of Technology, Kanpur, India}

\author{Patrick Rebentrost}\email{cqtfpr@nus.edu.sg}
\affiliation{Centre for Quantum Technologies, National University of Singapore, Singapore 117543}

\begin{abstract} 
Restricted Boltzmann Machines (RBMs) are widely used probabilistic undirected graphical models with visible and latent nodes, playing an important role in statistics and machine learning. The task of structure learning for RBMs involves inferring the underlying graph by using samples from the visible nodes. Specifically, learning the two-hop neighbors of each visible node allows for the inference of the graph structure. Prior research has addressed the structure learning problem for specific classes of RBMs, namely ferromagnetic and locally consistent RBMs. In this paper, we extend the scope to the quantum computing domain and propose corresponding quantum algorithms for this problem. Our study demonstrates that the proposed quantum algorithms yield a polynomial speedup compared to the classical algorithms for learning the structure of these two classes of RBMs.
\end{abstract}

\maketitle 

\section {Introduction}
Graphical models are widely used in probability theory and machine learning to describe the dependence structure among random variables. Various algorithms have been proposed for learning graphical models \cite{1054142, NIPS2016_861dc9bd, 10.1214/09-AOS691}. Models with latent (hidden) variables capture more complex dependencies compared to fully-visible models. Learning latent variable models using samples of visible variables is a major challenge in this field.

A Restricted Boltzmann Machine (RBM) is a two-layer network consisting of a visible layer and a hidden layer, where variables within each layer are not connected. It can be represented as a weighted bipartite graph connecting visible nodes and hidden nodes.
RBMs find applications in feature extraction and dimensionality reduction across various domains \cite{salakhutdinov2007restricted, larochelle2008classification}. However, learning general RBMs and related problems have been proven to be difficult \cite{long2010restricted}. Fortunately, certain classes of RBMs exhibit properties that can be efficiently learned.
One such property is the two-hop neighborhood of visible nodes, which refers to the visible nodes connected to a specific node through a single hidden node. Learning this property is an example of \textit{structure learning}, and knowing the two-hop neighborhood helps the learning of the marginal distribution of the visible nodes of the RBM. 

Bresler \textit{et al.} \cite{bresler2019learning} proposed a classical greedy algorithm based on influence maximization for learning the two-hop neighbors of a visible node for ferromagnetic RBMs.
In ferromagnetic RBMs, pairwise interactions between nodes and external fields are non-negative. 
The algorithm, based on the GHS (Griffiths-Hurst-Sherman) inequality, has a nearly quadratic runtime and logarithmic sample complexity with respect to the number of visible nodes. The runtime and sample complexity depend exponentially on the maximum degree, which is nearly optimal.
Additionally, Goel \cite{goel2019learning} extended the results to locally consistent RBMs, where pairwise interactions associated with each latent node have the same sign but arbitrary external fields are allowed. The proposed classical greedy algorithm for learning two-hop neighbors is based on maximizing conditional covariance, relying on the FKG (Fortuin–Kasteleyn–Ginibre) inequality. The runtime and sample complexity with respect to the number of visible nodes are the same as in \cite{bresler2019learning}, but the dependency on the upper bound strength is doubly exponential.

Quantum algorithms offer potential speed-ups over classical algorithms for certain problems \cite{grover1997quantum, durr1996quantum, HHL09}. Many quantum machine learning algorithms are based on amplitude amplification and estimation \cite{brassard2002quantum}, which can achieve quadratic speedups in some parameters while potentially slowing down others. Quantum learning of graphical models such as factor graphs was considered in \cite{gao2018quantum}. Quantum algorithms for learning generalized linear models, Sparsitron, and Ising models have been studied \cite{Rebentrost2021Hedge}. Quantum structure learning of MRFs has also been explored \cite{zhao2021quantum}. 
Quantum computation holds the promise of more efficient structure learning of RBMs and MRFs, which is of both theoretical and practical interest.

In this paper, we present quantum algorithms for structure learning of the underlying graphs of ferromagnetic RBMs and locally consistent RBMs with arbitrary external fields. The quantum algorithms are based on the classical algorithms in \cite{bresler2019learning} and \cite{goel2019learning} for non-degenerate RBMs with bounded two-hop degrees. We demonstrate that these quantum algorithms provide polynomial speed-ups over the classical counterparts in terms of sample dimensionality. Informally, our results are as follows.
\begin{theorem}[Informal version of Theorem \ref{th4} and Theorem \ref{th6}]
\label{informal_th1}\label{informal_th2}
Consider a ferromagnetic RBM (locally-consistent RBM) of $n$ visible nodes where pairwise interactions and external fields are upper and lower-bounded. There exists a quantum algorithm that learns the two-hop neighborhood for a single visible node with high probability in $\mathcal{\tilde{O}}(\sqrt{n})$ time and sample complexity close to the theoretical lower bound in $\supseteq \Omega(\log{n})$.
\end{theorem} 

The remainder of the paper is organized as follows.  
In Sec.~\ref{Peliminary}, we introduce the notations used in this paper and provide an overview of RBM models. In Sec.~\ref{Sec_ferromagnetic_RBM} we briefly review the classical greedy algorithm introduced in \cite{bresler2019learning} and present the quantum version. In Sec.~\ref{sec_local_consistent_RBM}, we provide a quantum algorithm for structure learning of locally consistent RBMs.

\section{Preliminaries}\label{Peliminary}

\textbf{Notations.}
Let $\mathbb Z_+$ denote the set of positive integers, $\mathbb R$ denote the set of real numbers, and  $[N]=\{1,2,\cdots,N\}$. For all sets $A \subset [N]$, define the indicator function $\mathbbm I_A$ as
\begin{eqnarray}
\mathbbm I_A(x):=\begin{cases}
1 \quad {\rm if} \quad  x\in A , \\ 
0 \quad {\rm otherwise.}
\end{cases}
\end{eqnarray}
Furthermore, let  $\sigma(x)={1}/\left(1+e^{-x}\right)$ denote the sigmoid function for $x\in \mathbb R$.

\begin{figure}[htpb!]
\includegraphics[width=0.8\linewidth]{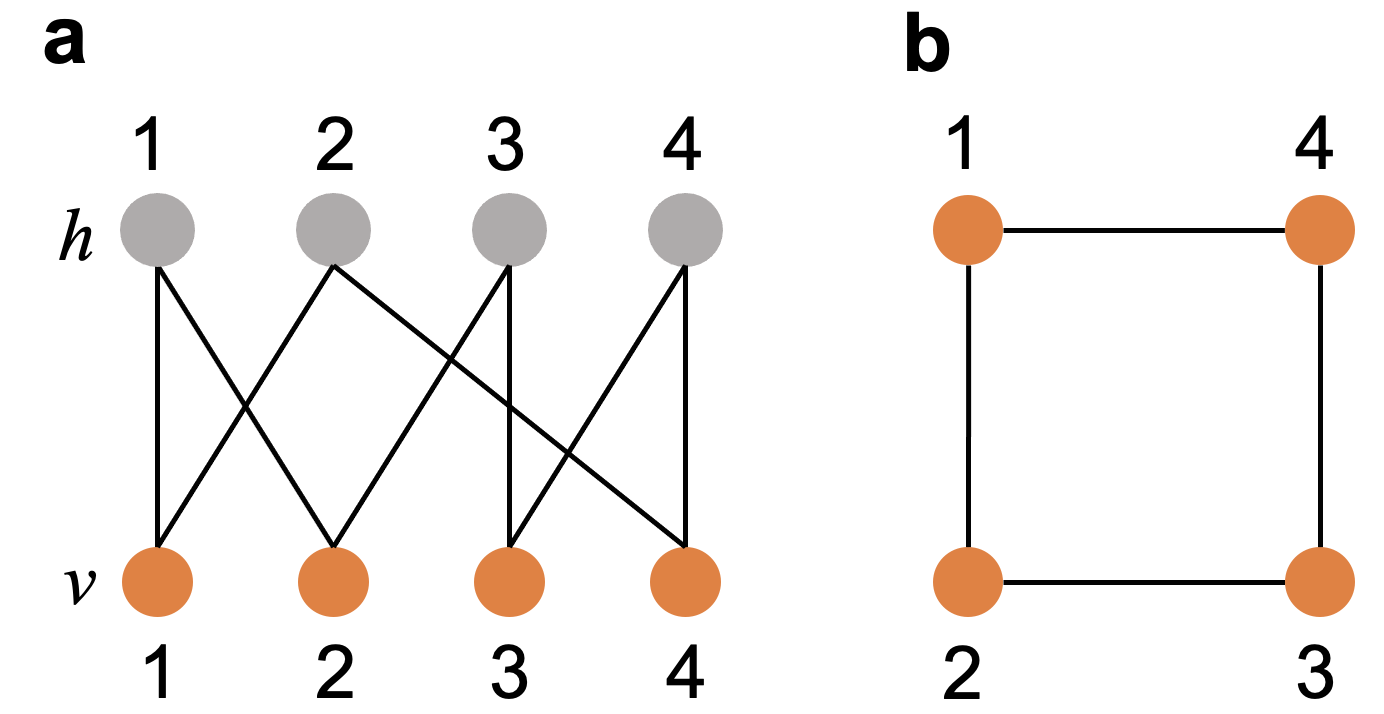}
  \caption{\textbf{a}. An example of an RBM with $4$ visible nodes (orange circles) and $4$ hidden nodes (grey circles). \textbf{b}. Underlying graph of the RBM in \textbf{a}.}\label{figure}
\end{figure}

 \textbf{Restricted Boltzmann Machines.}
Restricted Boltzmann Machines are a widely used class of graphical models with latent variables. An RBM consists of two layers, a visible layer, and a hidden layer. Nodes within the same layer are not connected. Thus, an RBM can be represented as a bipartite graph. Figure \ref{figure} (a) illustrates an example of an RBM with $4$ visible nodes and $4$ hidden nodes.
 Given an RBM with $n$ visible (observed) nodes $\{X_i\}_{i\in [n]}$ and $m$ latent nodes  $\{Y_k\}_{k\in [m]}$, the probability distribution for any configuration $x\in \{\pm1\}^n$ and $y\in \{\pm1\}^m$ is defined as
\begin{eqnarray}
P(X=x,Y=y)=\frac{1}{Z} \exp (x^{\mathsf{T}}Jy+f^{\mathsf{T}}{x} +g^{\mathsf{T}}y),
\end{eqnarray}
where $Z$ is the partition function, $f\in \mathbb{R}^{n}$ and $g\in \mathbb{R}^{m}$ are external fields, and $J\in \mathbb{R}^{n}\times \mathbb{R}^{m}$ is the interaction matrix.
An RBM with fixed external fields $f$ and $g$, and interaction matrix $J$ is denoted as RBM $(J,f,g)$.

Learning an RBM involves determining the optimal parameters $J, f$, and $g$.
In this paper, our focus is on the underlying structure learning of an RBM.
One approach to learning the structure of an RBM is to learn all two-hop neighbors of each visible node. The definition of two-hop neighbors of a visible node is provided below.

\begin{defn}[Two-hop Neighborhood] Let $i \in [n]$ be a visible node for a fixed 
RBM $(J, f, g)$. The two-hop neighborhood of $i$, denoted as $\mathcal N_2(i)$, is defined as the smallest set of visible nodes  $S \subset [n] \setminus \{i\}$ such that conditioned on $X_S$,  $X_i$ is conditionally independent of  $X_j$ for all visible nodes $j \in [n]\setminus (S \cup \{i\})$. The two-hop degree of the RBM is defined as  $ d_2=\max_{i \in[n]}\{|\mathcal N_2(i)|\}$.
\end{defn}

As mentioned in \cite{goel2019learning}, $\mathcal N_2(i)$ is always a subset of the graph-theoretic two-hop neighborhood, which is the smallest set $S$ such that vertex $i$ is separated from the other observable nodes in the graphical structure of the RBM.  
However, it may be a strict subset. By learning the two-hop neighbors of each visible node, we can reconstruct the underlying graph of the RBM. The underlying graph, denoted as $G$, represents the visible nodes as vertices, and an edge between two visible nodes $i$ and $j$ indicates that they are two-hop neighbors in the corresponding RBM. For example, in Fig. \ref{figure}, \textbf{b} is the underlying graph of the RBM in \textbf{a}. We observe that visible nodes $1$ and $2$ are two-hop neighbors in \textbf{a}, and correspondingly, there is an edge between these two nodes in the underlying graph \textbf{b}.

In order to learn the two-hop structure of an RBM, it is necessary to establish lower and upper bounds on the weights. A lower bound is needed to determine if there is an edge present in the RBM. If the interaction strength is weaker than the lower bound, we treat it as a non-edge.
On the other hand, an upper bound is required to ensure that the distribution of any variable is not close to a deterministic value.  This is a standard assumption in the literature on learning Ising models \cite{bresler2019learning,goel2019learning}. We consider $(\alpha, \beta)$ non-degenerate RBMs defined as follows. 
\begin{defn}
An RBM  is said to be $(\alpha,\beta)$-non-degenerate if it satisfies the following conditions: 
\begin{itemize}
\item for every $i\in [n], j\in[m],$ if $|J_{ij}|\neq 0$, we have $|J_{ij}|\geq \alpha.$ 
\item for every $i\in [n]$, we have $\sum_j|J_{ij}|+|f_i|\leq \beta$. 
\item for every $j\in [m]$, we have $\sum_i|J_{ij}|+|g_j|\leq \beta$. 
\end{itemize}
\end{defn}

\textbf{Classical and quantum computational model.} 
We assume standard classical and quantum computational models \cite{nielsen2010quantum}. 
In addition, a classical and corresponding quantum arithmetic model is assumed, which allows us to ignore issues arising from the fixed-point representation of real numbers and count the basic arithmetic operations as constant time operations.

\textbf{Quantum data input.}
Let us assume that we have quantum access to $M$ samples from a distribution of an RBM. We define the quantum data input as follows:
\begin{defn}[Quantum data input]\label{defQuantumSamples}
Let $\{x^{i} \in \{\pm 1\}^n\}_{i\in [M]}$ be $M$ samples of an RBM with $n$ visible nodes.
We say that we have quantum query access to the samples if for each sample $i\in[M]$ we have access to a quantum circuit that performs
\be
\ket j \ket {\bar 0} \to \ket j \ket{x^ {i}_j},
\ee
with  $\Ord{\log n}$ qubits in constant time, where $j \in [n]$ and $x^ {i}_j$ is the configuration of node $j$ in the $i$-th sample.
\end{defn}
Usually, this reversible operation is defined via the bit-wise modular addition, while here we only specify the action on the computational initial state $\ket {\bar 0}$.

\textbf{Quantum minimum finding.}
In this paper, we utilize a quantum subroutine known as quantum minimum finding \cite{durr1996quantum}. This subroutine provides an almost quadratic speedup compared to its classical counterpart.

\begin{theorem}[Quantum minimum finding \cite{durr1996quantum}]\label{quantum_max_finding}
Given quantum access to a comparison oracle for elements of a data set of size $N$, we can find the minimum element  with success probability $1-
\rho $ with $\Ord{\sqrt{N}\log(1/\rho)}$ queries and $\widetilde{O}(\sqrt{N}\log(1/\rho))$ quantum gates.
\end{theorem}
The minimum finding algorithm can be used to find the maximum element as well by changing the oracles accordingly.

\section{Structure Learning for ferromagnetic RBMs}\label{Sec_ferromagnetic_RBM}
In Ref.~\cite{bresler2019learning}, an efficient greedy algorithm based on influence maximization is described for learning the two-hop neighbors of any visible node in a ferromagnetic RBM. This algorithm allows for the recovery of the two-hop structure of the ferromagnetic RBM.
In this section, we first introduce the definition of the influence function. We then review the classical greedy algorithm presented in Ref.~\cite{bresler2019learning}. Finally, we propose a quantum version of the classical algorithm and demonstrate its polynomial speedup over the classical counterpart. 

The definition of a ferromagnetic RBM is provided below.
\begin{defn}[Ferromagnetic RBMs]
A ferromagnetic RBM is an RBM $(J,f,g)$ in which the pairwise interactions and external fields are all non-negative i.e., $J_{ij} \geq 0$, $f_i\geq 0$, and $g_j \geq 0$, $ \forall i \in [n], j \in [m]$.
\end{defn}
There are various measures to quantify correlations between variables, and one of them is the expected "magnetization" of a node when certain other nodes are fixed to $+1$. The formal definition is as follows.

\begin{defn}[Discrete Influence Function]
Given a visible node $u\in [n]$ and a subset $S\subset [n]\setminus \{u\}$  of a ferromagnetic RBM, let $s=\vert S \vert,$ the discrete influence function is defined as
\begin{eqnarray}
{I}_u(S) := \mathbb{E}\left [X_u| X_S=\{1\}^{s}\right].
\end{eqnarray}
\end{defn}
The discrete influence function is a monotone submodular function for any visible node $u \in [n]$ which has been proven using the Griffiths-Hurst-Sherman inequality in \cite{bresler2019learning}. It has also been shown that any set $S'$ that is close enough to the maximizer of $I_u(S')$ must contain the two-hop neighbors of $u$. 

Given $M$ samples from  a ferromagnetic RBM, the empirical discrete influence function is defined as follows
\begin{eqnarray}
\widehat{I}_u(S) := \widehat{\mathbb{E}}\left[X_u| X_S=\{1\}^{s}\right],\label{eq_influence_function}
\end{eqnarray}
where $\hat{\mathbb{E}}$ denotes the empirical expectation. 
Expanding the above equation yields: 
\begin{eqnarray}
\widehat{I}_u(S) &=&
\sum_{x_u\in \{\pm 1\}}x_u \widehat{P}(X_u=x_u \vert X_S=\{1\}^{s}) \nonumber\\
&=& \frac{2 \widehat{P}(X_{S\cup \{u \} }=\{1\}^{s+1})}{  \widehat{P}( X_S=\{1\}^{s}) } -1,
\label{influence_expand}
\end{eqnarray}
where the second line is obtained by using the Bayes' rule and the fact that $\sum_{x_u\in \{\pm 1\}}\widehat{P}(x_u \vert X_S=\{1\}^{s})=1. $
The empirical probability for $M$ samples  can be obtained by
\begin{eqnarray}
\widehat{P}(X=x) =\frac{1}{M}\textstyle\sum_{i=1}^M \mathbbm{1}_{\{X^{i}=x\}}.
\end{eqnarray} 

\textbf{Classical greedy algorithm.}
Given a number of samples from a ferromagnetic RBM, the two-hop neighbors of each visible node can be found from the maximizer of the empirical influence function \cite{bresler2019learning}. For a visible node $u\in[n]$, a visible node subset $S$ (excluding $u$), and a visible node $j$ that is neither $u$ nor part of $S$, it has been demonstrated that if $j$ is not a two-hop neighbor of $u$, the difference  ${I}_u(S\cup j)-{I}_u(S)$ equals zero. Conversely, if $j$ is a two-hop neighbor of $u$, we have ${I}_u(S\cup j)-{I}_u(S)\geq {2\eta}$, where the threshold $\eta$ will be defined later. Based on this insight, the authors propose a greedy algorithm, as outlined in Algorithm \ref{algo:greedy}, to identify the two-hop neighbors of each visible node. The algorithm's performance, including the required number of samples and the associated time complexity, has been analyzed in Ref.~\cite{bresler2019learning}.

\begin{theorem}[Theorem 6.1 in Ref~\cite{bresler2019learning}]\label{theorem_rbm_inf}
Given $M$ samples $\{x^{i} \in \{\pm1\}^n\}_{i\in[M]}$ of the visible nodes of a ferromagnetic RBM which is $(\alpha,\beta)$-non-degenerate, and has two-hop degree $d_2$. Let $\eta = \alpha^2 \sigma (-2\beta)(1-\tanh(\beta))^2$, $k=d_2\log(4/\eta)$,  for $\delta>0$, as long as 
\begin{eqnarray}
    M\geq 2^{2k+3}(d_2/\eta)^2(\log(n)+k\log(en/k))\log(4/\delta), \label{equ_sample_FRBM_classical}
\end{eqnarray}
 then for every visible node $u \in [n]$, the Algorithm \ref{algo:greedy} returns the set $\mathcal N_2(u)$ with probability $1-\delta$ in a time complexity of $\mathcal{O}(Mkn)$.
\end{theorem}

The total run time for all $n$ visible nodes is therefore $\mathcal{O}(Mkn^2)$. 
It is important to note that the number of iterations $k$ depends on the two-hop degree and the upper and lower bounds on the strengths of the RBM.

\begin{algorithm} 
\caption{\textsc{ClassicalGreedyFrbmNbhd}($u$)}
\label{algo:greedy}
\begin{algorithmic}[1]
\Require{ Visible node $u \in [n]$, threshold $\eta >0$, samples $ \{ x^{i} \in \{\pm1\}^n \}_{i \in [M]}$.}
\State Set $S_0 \gets \varnothing$.
\For {$t = 0,\cdots,k-1$}
\State{$j_{t+1} \gets \arg \max_{j\in [n]} \widehat{I}_u(S_t \cup \{j \})$.}\label{classical_ferro_step_min_finding}
\State $S_{t+1} \gets S_t \cup \{j_{t+1} \}$.
\EndFor
\State ${\hat{\mathcal N}}_2(u) \gets \{ j\in S_k : \widehat{I} _u(S_k) - \widehat{I}(S_k \setminus \{j\}) \geq \eta \}$.
\Ensure {$\hat{{\mathcal N}}_2(u)$}.
\end{algorithmic}\label{algGreedy}
\end{algorithm} 

\textbf{Quantum greedy algorithm.}
We then propose a quantum version of Algorithm \ref{algo:greedy} to learn the two-hop neighborhood of any visible node in a ferromagnetic RBM,  assuming we have quantum access to $M$ samples of the RBM. The key idea behind the quantum algorithm is to replace the step \ref{classical_ferro_step_min_finding} of Algorithm \ref{algo:greedy} with a quantum subroutine of maximum finding \cite{durr1996quantum}. This quantum subroutine is expected to provide a quadratic speed-up compared to the classical setting in step \ref{classical_ferro_step_min_finding}.

Firstly, we explore the quantum representation of the empirical discrete influence function.
Given quantum data input from Definition~\ref{defQuantumSamples}, we can prepare a quantum state that encodes the value of the empirical discrete influence $\widehat{I}_u(S\cup \{j\})$ as defined in Eq.~(\ref{eq_influence_function}). For the node subset $S$, we begin by identifying the samples where the configuration $x_S=\{1\}^s$ and storing the corresponding sample indices classically.

\begin{lemma}\label{lemma_FRBM_index_X_S_equal_to_1}
Given access to $M$ samples of a ferromagnetic RBM, for a given set $S\in[n]$, we can identity the samples where $x_S=\{1\}^s$ and record the corresponding sample index in a set 
\begin{eqnarray}
M_S:=\{i~\vert~ i\in[M], x_S^{i}=\{1\}^s\},
\end{eqnarray}
with $\Ord{Ms}$ queries.
\end{lemma}
\begin{proof}
Querying the $M$ samples for the set $S$ requires $\Ord{M s}$ queries. Hence, the sample indices where $x_S=\{1\}^s$  can be found and stored in the set $M_S$ in $\Ord{Ms}$ time.
\end{proof}

Since Eq.~(\ref{eq_influence_function}) can be expanded as Eq.~(\ref{influence_expand}), the empirical discrete influence $\widehat{I}_u(S\cup \{j\})$ can be obtained by using the probabilities ${ \widehat{P}(X_{S\cup \{u,j \} }=\{1\}^{s+2})}$ and ${\widehat{P}( X_{S\cup \{j\}}=\{1\}^{s+1})}$. We now show that the probabilities can be encoded into a quantum state. 
\begin{lemma}\label{lemma_rbm_prob}
Let us be given quantum query access to $M$ samples of the observable nodes of a ferromagnetic RBM according to Definition \ref{defQuantumSamples}. Let a subset $S \subset [n]$ with $\vert S \vert =s \leq k$ and $j \in [n]\setminus S$. Assuming we have the sample index set $M_S$ as defined in Lemma \ref{lemma_FRBM_index_X_S_equal_to_1}, there exists a quantum circuit that performs 
\begin{eqnarray}
 \ket j \ket {\bar 0} \to  \ket j  \ket{\widehat{P}\left(X_{S \cup \{j\}}=\{1\}^{s+1} \right) },\label{equ_probability_preparation}
\end{eqnarray}
with $\Ord{M}$ quantum queries and $\Ord{M}$ run time. 
\end{lemma}
\begin{proof}
The proof details are provided in Appendix \ref{appendix_A}.
\end{proof}

Similarly, by Lemma \ref{lemma_FRBM_index_X_S_equal_to_1}, we can find the set  $M_{S\cup \{u\}}$ which stores the sample indices where the configuration of each visible node in the set formed by combining $S$ with the node $u$ is equal to one. We can then prepare a quantum state representing the probability where the configuration of each node in $S \cup {j,u}$ is equal to one. 
Assuming we already have sets $M_S$ and $M_{S\cup \{u\}}$, we demonstrate that the empirical influence $\widehat{I}_u(S \cup \{j\})$ can be prepared in a quantum state. 
\begin{lemma} \label{lemmaInfluenceQuantum}
Let us be given quantum query access to $M$ samples of a ferromagnetic RBM according to Definition \ref{defQuantumSamples}.  Let $d_2$ be the two-hop degree and  $\eta, k  $ be as in Theorem \ref{theorem_rbm_inf}. Given a node $u\in[n]$, a subset $S \subset [n]\setminus\{u\}$ with $\vert S \vert \leq k$, and the sets $M_S$ and $M_{S\cup \{u\}}$ from Lemma \ref{lemma_FRBM_index_X_S_equal_to_1},  for $j\in [n]\setminus (S\cup \{u\})$,  there exists a unitary operator $U_{\rm inf}$ which performs
\be
 \ket j \ket {\bar 0} \to  \ket j \ket{\widehat{I}_u(S \cup \{j\})},\label{equ_main_influence_preparation}
\ee
where $\widehat{I}_u(S \cup \{j\})$ is the empirical influence function defined in Eq.(\ref{eq_influence_function}).
The circuit requires $\Ord{M}$  quantum queries and has a run time of $\Ord{M}$.
\end{lemma}

\begin{proof}
We can prepare a quantum state that encodes the empirical influence ${\widehat{I}_u(S \cup\{j\})}$ by following these steps. 
First, using Lemma \ref{lemma_rbm_prob}, for a visible node $j\in [n]\setminus (S\cup \{u\})$,  we can prepare state 
\begin{eqnarray}
  \ket j\ket{\widehat{P}\left(X_{S \cup\{j\}}=\{1\}^{s+1} \right) }
     \ket{\widehat{P}\left(X_{S \cup \{u,j\}}=\{1\}^{s+2}\right) }.~~
     \label{eq_main_probability_FRBM}
\end{eqnarray}
By calculating the influence based on these probabilities using Eq.~(\ref{influence_expand}), we can prepare the desired influence ${\widehat{I}_u(S \cup \{j\})}$ in a quantum state. Finally, we undo the unitary operator used in the state preparation process, resulting in the state given by Eq.~(\ref{equ_main_influence_preparation}).

Regarding the run time, according to Lemma \ref{lemma_rbm_prob}, preparing the state in Eq.~(\ref{eq_main_probability_FRBM}) requires $\Ord{M}$ queries and has a run time of $\Ord{M}$. Therefore, the total number of quantum queries required is $\Ord{M}$, and the total run time is $\Ord{M}$.
\end{proof}

\textbf{Main result.} 
We present the quantum version of the greedy algorithm (Algorithm \ref{quantum_algo:greedy}) for identifying the two-hop neighbors of a visible node $u$ using quantum access to samples from a ferromagnetic RBM. In this algorithm, we leverage the computation of the empirical influence  $\widehat{I}_u(S\cup \{ j \})$, which is stored as a bit string in a register of qubits. This allows us to operate on the superposition of all visible nodes $j\in[n]\setminus (S\cup \{u\})$. 
The key step in the quantum greedy algorithm is to apply the quantum minimum finding algorithm to determine the maximum value of the empirical influence $\widehat{I}_u(S\cup \{ j \})$ among all possible nodes $j$. By using a threshold $\eta>0$, we can then identify the neighbors of node $u$ based on whether the empirical influence surpasses this threshold.

\begin{algorithm}
\caption{\textsc{QuantumGreedyFrbmNbhd}($u$)}
\label{quantum_algo:greedy}
\begin{algorithmic}[1]
\Require{Visible node $u \in [n]$, threshold $\eta >0$, $\delta >0$,  quantum query access to samples $\{ x^{i} \in \{\pm1\}^n\}_{i \in [M] }.$}
\State Set $S_0 \gets \varnothing$.

\For {$t = 0,\cdots,k-1$}
\State{$j_{t+1} \gets \arg \max_{j\in [n]} \widehat{I}_u(S_t \cup \{j \})$ by using Quantum maximum finding in Lemma \ref{quantum_max_finding}  with success probability $1-\frac{\delta}{2k}$, this procedure requires applying $U_{\rm inf}$ in Lemma \ref{lemmaInfluenceQuantum} a number of times.}\label{algorithm_step_3_quantumm_FRBM}
\State{$S_{t+1} \gets S_t \cup \{j_{t+1} \}$.}
\EndFor
\State $\hat{\mathcal N}_2(u) \gets \{ j\in S_k : \widehat{I}_u(S_k) - \widehat{I}(S_k \setminus \{j\}) \geq \eta \}$.
\Ensure {$\hat{\mathcal N}_2(u)$}
\end{algorithmic}
\end{algorithm}

\begin{theorem} 
\label{th4}
Let us be given quantum access to samples $\{x^{i} \in \{\pm1\}  : i\in [M] \}$ from the observable distribution of an $(\alpha,\beta)$ non-degenerate ferromagnetic RBM with two-hop degree $d_2$, for any node $u$. Let $\eta, k$ be as in Theorem \ref{theorem_rbm_inf}. If the number of samples satisfies 
\begin{eqnarray}
 M\geq 2^{2k+3}(d_2/\eta)^2\left(\log(n)+k\log(en/k)\right)\log(8/\delta), \label{equ_sample_FRBM_quantum}
\end{eqnarray}
 for every visible node $u$, Algorithm \ref{quantum_algo:greedy} returns $\mathcal N_2(u)$ with success probability at least $1-\delta$ in time $\Ord{Mk^2 +Mk\sqrt{n} \log({k}/{\delta})}.$ 
\end{theorem}
\begin{proof}
The number of samples required has been proved in Ref.~\cite{bresler2019learning}, notice that the sample required is slightly different between the classical algorithm and the quantum version, due to the different success probability. Here we analyze the query complexity. Step  \ref{algorithm_step_3_quantumm_FRBM} in the for loop is the most time-consuming step. By using Lemma \ref{lemma_FRBM_index_X_S_equal_to_1}, Lemma \ref{lemmaInfluenceQuantum}  and Lemma \ref{quantum_max_finding}, the number of queries required in step \ref{algorithm_step_3_quantumm_FRBM} is  $\Ord{Mk+M\sqrt{n} \log ({k}/{\delta})}.$ Thus,  the total run-time of the algorithm is $\Ord{Mk^2+Mk \sqrt{n}\log({k}/{\delta})}$ for $k$ iterations. 

We now analyze the success probability. In our setting, the success probability for every quantum maximum finding is $1-{\delta}/{(2k)}$. For $k$ loop, the success probability is $(1-{\delta}/{(2k)})^k \geq 1- {\delta}/{2}$. Combining this with the success probability of the original classical algorithm, which we set as $1-\frac{\delta}{2}$, the total success probability is at least $1-\delta$ by the union bound.
\end{proof}

The total run time is then $\Ord{Mkn\left(k+ \sqrt{n} \log ({k}/{\delta})\right)}$ for finding the two-hop neighbors for all $n$ visible nodes.  With this information, we can obtain the structure of the underlying graph of the RBM.  

\section{Structure learning for locally consistent RBM}\label{sec_local_consistent_RBM}

In Ref \cite{goel2019learning}, a classical greedy algorithm was introduced for learning the two-hop neighbors of any visible node in locally consistent RBMs with arbitrary external fields. The algorithm maximizes the conditional covariance between pairs of visible nodes, based on the observation that the covariance is positive and bounded away from $0$ for pairs of visible nodes that are two-hop neighbors (connect to a common hidden node). This property also holds for the ferromagnetic Ising model with arbitrary external field \cite{bresler2015efficiently}.
In this section, we will review the classical greedy algorithm, then propose a quantum version of the algorithm, and demonstrate its improved efficiency.  

We start by defining a locally consistent RBM, which is an RBM where the interaction weights between each hidden node and visible nodes are either all non-negative or non-positive. Formally,  a locally consistent RBM can be defined as follows:
\begin{defn}[Locally Consistent RBMs]
An RBM $(J,f,g)$ is locally consistent if for each  $j\in [m]$, we have $J_{ij}\geq 0$ for all $i \in [n]$, or $J_{ij}\leq 0$ for all $i \in [n].$
\end{defn} 

The conditional covariance is defined as follows. 
\begin{defn}[Conditional covariance]
The conditional covariance for visible node $u,v\in [n]$ and a subset $S \subseteq  [n] \setminus  \{u,v\}$   is defined as
\begin{eqnarray}
 \text{Cov}(u,v|x_S)
:=  \mathbb{E}[X_uX_v|x_S]
- \mathbb{E}[X_u|x_S ]\mathbb{E}[X_v|x_S],\nonumber
\end{eqnarray}
where $x_S$ is a shorthand notation for $X_S=x_S.$
The average conditional covariance is then defined as
\begin{eqnarray}
 \text{Cov}^{\text{avg}}(u,v|S) := \mathbb{E}_{x_S}[\text{Cov}(u,v|x_S) ].  
\end{eqnarray}
\end{defn}

Given a number of  samples of visible nodes of an RBM, the empirical average conditional covariance is defined as
\begin{eqnarray}
 \widehat{\text{Cov}}^{\text{avg}}(u,v \vert S)
&=& \widehat{\mathbb{E}}_{x_S}[ \widehat{\text{Cov}}(u,v|x_S)], \label{empi_ave_cov} 
\end{eqnarray}
where $\mathbb{E}$ is the empirical expectation and $\widehat{\text{Cov}}(u,v|x_S)$ is the empirical conditional covariance.\\

\textbf{Classical greedy algorithm.}
In Ref~\cite{goel2019learning}, it has been shown that for any visible node $u\in[n]$ and a visible node subset $S\in[n]$ that does not contain $u$, if a visible node $v\in[n]$ is neither equal to $u$ nor contained in $S$ 
is not a two-hop neighbor of $u$, the conditional covariance $\text{Cov}(u,v|x_S)$ is equal to zero. Conversely, if $v$ is a two-hop neighbor of $u$, we have $\text{Cov}(u,v|x_S)\geq 2\tau$,  where $\tau$ is a function of $\alpha$ and $\beta$ which will be provided later. 
Based on these observations, the author proposed a greedy algorithm (Algorithm~\ref{alg_rbm_cov}) to learn the two-hop neighbors of each node by maximizing the empirical average conditional covariance.

\begin{algorithm}
\caption{\textsc{ClassicalGreedyLrbmNbhd}($u$) } 
\label{alg_rbm_cov}
\begin{algorithmic}[1]
\Require{Visible node $u$, samples $\{x^{i} \in \{\pm1\}^n \}_{i \in [H]}$, threshold $\tau>0$.}
\State Set $S:=\emptyset $. 
\State{Let
$ i^*, \text{val}^*  = \textstyle\arg\max_v\widehat{\text{Cov}}^{\text{avg}}(u,v\vert S), \textstyle\max_v\widehat{\text{Cov}}^{\text{avg}}(u,v\vert S)$}\label{algorithm_classical_LRBM_step_2}
\If {$\text{val}^*\geq \tau$} 
        \State $ S=S\cup \{i^*\}$.
\Else  
\State Go to step 9.
\EndIf 
\State Go to step 2.
\State Pruning step: For each $v\in S$, if $\widehat{\text{Cov}}^{\text{avg}}(u,v\vert S) <\tau$, remove $v$.
\Ensure{$ S\to \hat{\mathcal{N}}_2(u)$.}
\end{algorithmic}
\end{algorithm}
The following Theorem gives the number of samples required and the run time of the algorithm.  
\begin{theorem}[Theorem 6 in Ref~\cite{ goel2019learning}]\label{theorem_classical_LRBM}
Given $H$ samples of visible nodes $\{x^{i}\in \{\pm 1\}^n\}_{i\in [H]}$ of an $(\alpha,\beta)$-nondegenerate locally consistent RBM, for $\delta =\frac{1}{2} e^{-2\beta}$,  $\tau= \alpha^2 \exp(-12\beta)$ and $T^*=\frac{8}{\tau^2}$, if 
\begin{eqnarray}
H \geq \Omega \left( \left( \log(1/ \zeta)+T^* \log n\right)\frac{2^{2T^*}}{\tau^2\delta^{2T^*}} \right),\label{eq_lrbm_sample}
\end{eqnarray}
 the two-hop neighbours $\mathcal N_2(u)$ of a visible $u$ can be obtained in time $O(HnT^*)$ using Algorithm \ref{alg_rbm_cov}  with success probability at least $1-\zeta.$
\end{theorem}

The proof of this theorem can be found in Ref.~\cite{goel2019learning}.\\

\textbf{Quantum greedy algorithm.}
Based on Algorithm \ref{alg_rbm_cov}, we propose a quantum version of the algorithm to learn the underlying graph of a locally consistent RBM with arbitrary external fields. We assume that we have quantum access to a number of samples from such an RBM, as defined in Definition \ref{defQuantumSamples}. The main idea is to utilize quantum techniques to prepare the empirical average conditional covariance $\widehat{\text{Cov}}^{\text{avg}}(u,v\vert S)$ in a quantum state with a superposition of all node $v$, and then use a quantum minimum finding algorithm to find the values $i^*$ and $\text{val}^*$ in step \ref{algorithm_classical_LRBM_step_2} of Algorithm \ref{alg_rbm_cov}. Our approach still involves a hybrid algorithm, combining classical and quantum computations.

For each node $u$ and a known set $S$, we first determine the unique configurations of the visible nodes on that subset $S$ in the $H$ samples.  For a subset $S\in [n]$, there are $2^{s}$ possible configurations of the visible nodes on that subset.
Let $L_S$ denote the set consisting of the unique configurations of $S$ in the $H$ samples. There are at most $\vert L_S \vert \leq \min \{H, 2^{s}\}$ unique configurations, where $s$ is the size of the subset $S$. Formally, we define the unique configuration set $L_S$ and the set $F(x_S^{l})$ which stores the sample indices in $[H]$ where $l\in [\vert L_S \vert]$ and the configuration of set $S$ is equal to $x_S^{l}$ in set $L_S$, as follows:
\begin{eqnarray}
L_S & := & {\rm Unique} \{ x^{i}_S : i \in [H]\},\nonumber\\
F(x_S^{l}) & := & \{i ~\vert~ i\in[H], ~x_S^{i} = x_S^{l} \},\label{definition_L_S_F}
\end{eqnarray} where ${\rm Unique} $ returns the set of unique elements. 
For example, ${\rm Unique} \{a,b,c,d,d,d,a\} = \{a,b,c,d\}$.

\begin{lemma}\label{unique_S_via_quantu_access}
Given quantum access to $H$ samples of an RBM as Definition \ref{defQuantumSamples}, for a subset $S\subset [n],$ we can obtain the unique configuration set $L_S$ and  sample index set $F(x_S^{l})_{l\in[\vert L_S \vert]}$ defined in Eq.~(\ref{definition_L_S_F}) with $\Ord{Hs}$ quantum queries and $\Ord{Hs}$ runtime. 
\end{lemma}
\begin{proof}
To obtain the unique configuration set $L_S$, we first query the quantum access to the $H$ samples as described in Definition \ref{defQuantumSamples}. This allows us to obtain the quantum state
\begin{eqnarray}
\ket{S}\ket{\bar{0}}\to \ket{S}\otimes_{i=1}^{H}\ket{x_S^{i}},
\end{eqnarray}
with $\mathcal{O}(Hs)$ queries. We can then measure the quantum state in the computational basis to obtain the classical representation of the configurations. By examining all $H$ samples, we can identify the unique configurations and construct the set $L_S$. This process takes $\mathcal{O}(Hs)$ time.

Therefore, we can obtain the unique configuration set $L_S$ and the sample index set $F(x_S^{l})_{l\in[\vert L_S \vert]}$ with $\mathcal{O}(Hs)$ quantum queries and $\mathcal{O}(Hs)$ runtime.
\end{proof}

Now we demonstrate how to compute the empirical average conditional covariance. Let $x_S^{l}$ denote the $l$-{th} configuration in set $L_S$ defined in Eq.~(\ref{definition_L_S_F}). 
The empirical average conditional covariance in Eq.~(\ref{empi_ave_cov}) can be expressed as 
\begin{eqnarray}
\widehat{\text{Cov}}^{\text{avg}}(u,v \vert S) = \frac{1}{H} \left( \sum_{i=1}^{H} z_{u,v}^{i}-  \sum_{l=1}^{\vert L_S\vert}   \frac{ a_{u,x_S^{l}}a_{v,x_S^{l}}}{ \vert {F}(x_S^{l}) \vert }\right), ~\label{main_eqcov}
\end{eqnarray}
where $z_{u,v}^{i} :=  x_{u}^{i}x_{v}^{i}$ represents the product of configurations of $u$ and $v$ in the $i$-th sample, and
\begin{eqnarray}
a_{j,x_S^{l}}  :=\textstyle\sum_{i\in F(x_S^{l})} x^{i}_j,~ \text{for}~ j=u,v\label{eq_a_j_X_S}
\end{eqnarray}
denote the sum of the configurations of $u$ ($v$) in the samples where the sample index is in $F(x_S^{l}).$
The complete derivation of Eq.~(\ref{main_eqcov}) is provided in Appendix \ref{appendix_cov}.

We now show that the empirical average conditional covariance in Eq~(\ref{main_eqcov}) can be represented in a quantum state. From Lemma \ref{unique_S_via_quantu_access}, we know that obtaining the unique configuration set requires  $\Ord{Hs}$ quantum queries and running time. 
As it only needs to be calculated once for set $S,$ we assume we have already gotten different configurations of the $H$ samples for set $S$. We proceed with the representation.

\begin{lemma}\label{lemma_quantum_covariance}
Given quantum query access to $H$ samples of an RBM as defined in Definition \ref{defQuantumSamples},  nodes $u,v\in[n]$, set $S \subset [n]\setminus\{u,v\}$,  the unique configuration set $L_S$ and the sample indexes subset $F(x_S^{l})$ as defined in Eq.~(\ref{definition_L_S_F}), there exists a unitary $U_{cov}$ which performs the transformation
\begin{eqnarray}
\ket{v}\ket{u}\ket{\bar{0}}  \rightarrow    \ket{v}\ket{u}\ket{\widehat{\text{Cov}}^{\text{avg}}(u,v\vert S)},\label{equlemma1}
\end{eqnarray}
in $\Ord{H}$ quantum queries and $\Ord{H}$ running time.
\end{lemma}
The detailed proof is provided in Appendix \ref{appendix_proof_lemma_cov}.\\

\textbf{Main result.} We introduce the quantum version of Algorithm \ref{alg_rbm_cov} in Algorithm \ref{alg_quantum_rbm_cov}, designed to identify the two-hop neighbors of a locally consistent RBM.
According to Lemma \ref{lemma_quantum_covariance}, we observe that the empirical average condition covariance $\widehat{\,\text{Cov}}^{\text{avg}}(u,v|S)$ can be encoded in  a quantum state with a supposition of all visible node $v$ which is neither  $u$ and nor contained  in set $S$. Then by utilizing the quantum minimum finding algorithm, we can determine a node among all nodes $v$ which maximizes the empirical average conditional covariance. By comparing this value with the threshold $\tau$, we can determine whether $v$ is a two-hop neighbor of $u$.

\begin{theorem}
\label{th6}
Given quantum access to $H$ samples of a locally consistent RBM according to Definition \ref{defQuantumSamples}, if the number of samples satisfies Eq.~(\ref{eq_lrbm_sample}), we can find the two-hop neighbors of any visible node $u$ with success probability at least $1-\zeta$, using  $\Ord{H T^{*2}+H\sqrt{n}T^* \log({T^*}/{\zeta})}$ runtime. 
\end{theorem}
\begin{proof}
We observe that the sample complexity remains the same as in Theorem \ref{theorem_classical_LRBM}.
Next, we analyze the query complexity and runtime of Algorithm \ref{alg_quantum_rbm_cov}. The most time-consuming steps are the loop steps 2-9, particularly steps 2 and 3. As  shown in Lemma \ref{unique_S_via_quantu_access}, step $2$ requires $\Ord{Hs}$ quantum queries and $\Ord{Hs}$ run time. For step $3$, 
as discussed earlier in Lemma~\ref{quantum_max_finding}, we can employ the  quantum minimum finding algorithm \cite{durr1996quantum} to find the maximum element $\text{val}^*$ and the corresponding visible node $i^*$  from the set of elements $\{\widehat{\,\text{Cov}}^{\text{avg}}(u,v|S)\, |\, v \in [n]/(S \cup \{u\})\,\}$ given as quantum states. In particular, $\text{val}^*$ and $i^*$ can be found by performing $\Ord{\sqrt{n}\log ({T^*}/{\zeta})}$ times of unitary $U_{cov}$ in Lemma \ref{lemma_quantum_covariance}  with success probability $1-\frac{\zeta}{2T^*}$. The  number of quantum query required is then $\Ord{H \sqrt{n}\log ({T^*}/{\zeta})}$ as each $U_{cov}$ requires $\Ord{H}$ quantum queries in step 3. Combining steps 2 and 3, 
the total number of quantum queries and runtime is
 $\Ord{Hs+H\sqrt{n} \log({T^*}/{\zeta})}.$  
  Over $T^*$ iterations, the runtime becomes $\Ord{HT^{*2} +  H\sqrt{n}T^*\log({T^*}/{\zeta})}\nonumber
$ as $s\leq T^*.$

We now turn to analyze the success probability. For each quantum maximum finding, the success probability is $1-{\zeta}/{(2T^*)}$, and over $T^*$ iterations, the success probability becomes $(1-{\zeta}/{(2T^*)})^L \geq 1- {\zeta}/{2}$.  Combining this with the success probability of the original classical algorithm, which we set as $1-{\zeta}/{2}$, this leads to a total success probability $(1-{\zeta}/{2})^2\geq 1-\zeta.$  
\end{proof}

\begin{algorithm}
\caption{\textsc{QuantumGreedyLrbmNbhd}($u$)} 
\label{alg_quantum_rbm_cov}
\begin{algorithmic}[1]
\Require{Visible node $u$, quantum query access to samples $\{ \,x^{i} \in \{\pm1\}^n \}_{i \in [H]} $, threshold $\tau$. }
\State Set $S:=\emptyset $. 
\State Find set $L_S$ and $F(x_S^l)$ for $l\in [\vert L_S \vert]$ defined in Eq.~(\ref{definition_L_S_F}).
\State {Find a $i^* \leftarrow \arg\max_v\widehat{\text{Cov}}^{\text{avg}}(u,v\vert S)$ by quantum maximum finding, via apply number of $U_{cov}$ in Lemma \ref{lemma_quantum_covariance} with success probability $1-\frac{\zeta}{2T^*}.$}
\If { $\widehat{\text{Cov}}^{\text{avg}}(u,i^*\vert S)\geq \tau$.} 
\State  $ S=S\cup \{i^*\}$.
\Else  
\State Go to step 9.
\EndIf 
\State Go to step 2.
\State Pruning step: For each $v\in S$, if $\widehat{\text{Cov}}^{\text{avg}}(u,v\vert S) <\tau$, remove $v$.
\Ensure{$S\to \hat{\mathcal{N}}_2(u)$.}
\end{algorithmic}
\end{algorithm}

Applying Algorithm \ref{alg_quantum_rbm_cov} to all $n$ visible nodes allows us to learn the underlying graph of the RBM. The total run time is then $\Ord{nHT^{*}\left(T^{*} +  \sqrt{n}\log({T^*}/{\zeta}\right)}
$.
\\

\section{Discussion and conclusion}\label{sec_discussion_conclusion}
In this work, we  have presented quantum algorithms for learning the two-hop neighbors of  ferromagnetic RBMs and locally consistent RBMs with arbitrary external fields.  Our quantum algorithms offer a polynomial speedup over their classical counterparts in terms of the number of visible nodes, while maintaining the same sample complexity. 
By exploiting quantum query access to the RBM samples, we can efficiently obtain the unique configuration set and compute the empirical average conditional covariance. This enables a speedup in the identification of visible nodes that have the highest covariance with a given node, indicating their two-hop neighbor relationship.

Once the structure of the underlying graph is obtained, further analysis and modeling can be performed. For example, we can apply existing algorithms such as the Sparsitron algorithm \cite{klivans2017learning} or GLMtron \cite{NIPS2011_30bb3825} to learn the parameters of the RBMs. These algorithms take advantage of the sparsity of the graph structure to achieve efficient parameter estimation.

Structure learning is a fundamental problem in machine learning, and our quantum algorithms offer the potential for speedup in other graph-based learning tasks as well. This theoretical research is inspired by numerous applications where learning the underlying structure of data is crucial, such as in social network analysis, biological network inference, and recommendation systems. 
Exploring provable learning using quantum algorithms shows  the prospects and limits of learning in a theoretical setting, which translates into insights for practical settings as well.

In conclusion, our work demonstrates the advantages of quantum computing in accelerating the learning of two-hop neighbors in RBMs. The quantum algorithms we have developed provide a promising avenue for structure learning tasks and open up new possibilities for efficient data analysis.  We anticipate further advancements in leveraging quantum techniques for graph-based learning and other related problems.

\section{Acknowledgements}
The author LZ was supported by the National Natural Science Foundation of China (No.12204386), and the Scientific and Technological Innovation Project (No. 2682023CX084).
Additionally, this research is supported by the National Research Foundation, Singapore under its CQT Bridging Grant. AA acknowledges an internship visit at CQT. 

\bibliographystyle{apsrev}
\bibliography{MRF}

\begin{thebibliography}{19}
\expandafter\ifx\csname natexlab\endcsname\relax\def\natexlab#1{#1}\fi
\expandafter\ifx\csname bibnamefont\endcsname\relax
  \def\bibnamefont#1{#1}\fi
\expandafter\ifx\csname bibfnamefont\endcsname\relax
  \def\bibfnamefont#1{#1}\fi
\expandafter\ifx\csname citenamefont\endcsname\relax
  \def\citenamefont#1{#1}\fi
\expandafter\ifx\csname url\endcsname\relax
  \def\url#1{\texttt{#1}}\fi
\expandafter\ifx\csname urlprefix\endcsname\relax\def\urlprefix{URL }\fi
\providecommand{\bibinfo}[2]{#2}
\providecommand{\eprint}[2][]{\url{#2}}

\bibitem[{\citenamefont{Chow and Liu}(1968)}]{1054142}
\bibinfo{author}{\bibfnamefont{C.}~\bibnamefont{Chow}} \bibnamefont{and}
  \bibinfo{author}{\bibfnamefont{C.}~\bibnamefont{Liu}}, \bibinfo{journal}{IEEE
  Transactions on Information Theory} \textbf{\bibinfo{volume}{14}},
  \bibinfo{pages}{462} (\bibinfo{year}{1968}).

\bibitem[{\citenamefont{Vuffray et~al.}(2016)\citenamefont{Vuffray, Misra,
  Lokhov, and Chertkov}}]{NIPS2016_861dc9bd}
\bibinfo{author}{\bibfnamefont{M.}~\bibnamefont{Vuffray}},
  \bibinfo{author}{\bibfnamefont{S.}~\bibnamefont{Misra}},
  \bibinfo{author}{\bibfnamefont{A.}~\bibnamefont{Lokhov}}, \bibnamefont{and}
  \bibinfo{author}{\bibfnamefont{M.}~\bibnamefont{Chertkov}}, in
  \emph{\bibinfo{booktitle}{Advances in Neural Information Processing Systems}}
  (\bibinfo{publisher}{Curran Associates, Inc.}, \bibinfo{year}{2016}),
  vol.~\bibinfo{volume}{29}.

\bibitem[{\citenamefont{Ravikumar et~al.}(2010)\citenamefont{Ravikumar,
  Wainwright, and Lafferty}}]{10.1214/09-AOS691}
\bibinfo{author}{\bibfnamefont{P.}~\bibnamefont{Ravikumar}},
  \bibinfo{author}{\bibfnamefont{M.~J.} \bibnamefont{Wainwright}},
  \bibnamefont{and} \bibinfo{author}{\bibfnamefont{J.~D.}
  \bibnamefont{Lafferty}}, \bibinfo{journal}{The Annals of Statistics}
  \textbf{\bibinfo{volume}{38}}, \bibinfo{pages}{1287 } (\bibinfo{year}{2010}).

\bibitem[{\citenamefont{Salakhutdinov et~al.}(2007)\citenamefont{Salakhutdinov,
  Mnih, and Hinton}}]{salakhutdinov2007restricted}
\bibinfo{author}{\bibfnamefont{R.}~\bibnamefont{Salakhutdinov}},
  \bibinfo{author}{\bibfnamefont{A.}~\bibnamefont{Mnih}}, \bibnamefont{and}
  \bibinfo{author}{\bibfnamefont{G.}~\bibnamefont{Hinton}}, in
  \emph{\bibinfo{booktitle}{Proceedings of the 24th international conference on
  Machine learning}} (\bibinfo{year}{2007}), pp. \bibinfo{pages}{791--798}.

\bibitem[{\citenamefont{Larochelle and
  Bengio}(2008)}]{larochelle2008classification}
\bibinfo{author}{\bibfnamefont{H.}~\bibnamefont{Larochelle}} \bibnamefont{and}
  \bibinfo{author}{\bibfnamefont{Y.}~\bibnamefont{Bengio}}, in
  \emph{\bibinfo{booktitle}{Proceedings of the 25th international conference on
  Machine learning}} (\bibinfo{year}{2008}), pp. \bibinfo{pages}{536--543}.

\bibitem[{\citenamefont{Long and Servedio}(2010)}]{long2010restricted}
\bibinfo{author}{\bibfnamefont{P.~M.} \bibnamefont{Long}} \bibnamefont{and}
  \bibinfo{author}{\bibfnamefont{R.~A.} \bibnamefont{Servedio}},
  \bibinfo{journal}{In Proceedings of the 27th International Conference on
  Machine Learning (ICML-10)} pp. \bibinfo{pages}{703--710}
  (\bibinfo{year}{2010}).

\bibitem[{\citenamefont{Bresler et~al.}(2019)\citenamefont{Bresler, Koehler,
  and Moitra}}]{bresler2019learning}
\bibinfo{author}{\bibfnamefont{G.}~\bibnamefont{Bresler}},
  \bibinfo{author}{\bibfnamefont{F.}~\bibnamefont{Koehler}}, \bibnamefont{and}
  \bibinfo{author}{\bibfnamefont{A.}~\bibnamefont{Moitra}}, in
  \emph{\bibinfo{booktitle}{Proceedings of the 51st Annual ACM SIGACT Symposium
  on Theory of Computing}} (\bibinfo{year}{2019}), pp.
  \bibinfo{pages}{828--839}.

\bibitem[{\citenamefont{Goel}(2020)}]{goel2019learning}
\bibinfo{author}{\bibfnamefont{S.}~\bibnamefont{Goel}}, in
  \emph{\bibinfo{booktitle}{Proceedings of the Twenty Third International
  Conference on Artificial Intelligence and Statistics}}, edited by
  \bibinfo{editor}{\bibfnamefont{S.}~\bibnamefont{Chiappa}} \bibnamefont{and}
  \bibinfo{editor}{\bibfnamefont{R.}~\bibnamefont{Calandra}}
  (\bibinfo{publisher}{PMLR}, \bibinfo{year}{2020}), vol. \bibinfo{volume}{108}
  of \emph{\bibinfo{series}{Proceedings of Machine Learning Research}}, pp.
  \bibinfo{pages}{3557--3566}.

\bibitem[{\citenamefont{Grover}(1997)}]{grover1997quantum}
\bibinfo{author}{\bibfnamefont{L.~K.} \bibnamefont{Grover}},
  \bibinfo{journal}{Physical Review Letters} \textbf{\bibinfo{volume}{79}},
  \bibinfo{pages}{325} (\bibinfo{year}{1997}).

\bibitem[{\citenamefont{D\"urr and H{\o}yer}(1996)}]{durr1996quantum}
\bibinfo{author}{\bibfnamefont{C.}~\bibnamefont{D\"urr}} \bibnamefont{and}
  \bibinfo{author}{\bibfnamefont{P.}~\bibnamefont{H{\o}yer}},
  \bibinfo{journal}{arXiv preprint quant-ph/9607014}  (\bibinfo{year}{1996}).

\bibitem[{\citenamefont{Harrow et~al.}(2009)\citenamefont{Harrow, Hassidim, and
  Lloyd}}]{HHL09}
\bibinfo{author}{\bibfnamefont{A.}~\bibnamefont{Harrow}},
  \bibinfo{author}{\bibfnamefont{A.}~\bibnamefont{Hassidim}}, \bibnamefont{and}
  \bibinfo{author}{\bibfnamefont{S.}~\bibnamefont{Lloyd}},
  \bibinfo{journal}{Physical Review Letters} \textbf{\bibinfo{volume}{103}}
  (\bibinfo{year}{2009}).

\bibitem[{\citenamefont{Brassard et~al.}(2002)\citenamefont{Brassard, Hoyer,
  Mosca, and Tapp}}]{brassard2002quantum}
\bibinfo{author}{\bibfnamefont{G.}~\bibnamefont{Brassard}},
  \bibinfo{author}{\bibfnamefont{P.}~\bibnamefont{Hoyer}},
  \bibinfo{author}{\bibfnamefont{M.}~\bibnamefont{Mosca}}, \bibnamefont{and}
  \bibinfo{author}{\bibfnamefont{A.}~\bibnamefont{Tapp}},
  \bibinfo{journal}{Contemporary Mathematics} \textbf{\bibinfo{volume}{305}},
  \bibinfo{pages}{53} (\bibinfo{year}{2002}).

\bibitem[{\citenamefont{Gao et~al.}(2018)\citenamefont{Gao, Zhang, and
  Duan}}]{gao2018quantum}
\bibinfo{author}{\bibfnamefont{X.}~\bibnamefont{Gao}},
  \bibinfo{author}{\bibfnamefont{Z.-Y.} \bibnamefont{Zhang}}, \bibnamefont{and}
  \bibinfo{author}{\bibfnamefont{L.-M.} \bibnamefont{Duan}},
  \bibinfo{journal}{Science advances} \textbf{\bibinfo{volume}{4}},
  \bibinfo{pages}{eaat9004} (\bibinfo{year}{2018}).

\bibitem[{\citenamefont{Rebentrost et~al.}(2021)\citenamefont{Rebentrost,
  Hamoudi, Ray, Wang, Yang, and Santha}}]{Rebentrost2021Hedge}
\bibinfo{author}{\bibfnamefont{P.}~\bibnamefont{Rebentrost}},
  \bibinfo{author}{\bibfnamefont{Y.}~\bibnamefont{Hamoudi}},
  \bibinfo{author}{\bibfnamefont{M.}~\bibnamefont{Ray}},
  \bibinfo{author}{\bibfnamefont{X.}~\bibnamefont{Wang}},
  \bibinfo{author}{\bibfnamefont{S.}~\bibnamefont{Yang}}, \bibnamefont{and}
  \bibinfo{author}{\bibfnamefont{M.}~\bibnamefont{Santha}},
  \bibinfo{journal}{Phys. Rev. A} \textbf{\bibinfo{volume}{103}},
  \bibinfo{pages}{012418} (\bibinfo{year}{2021}).

\bibitem[{\citenamefont{Zhao et~al.}(2021)\citenamefont{Zhao, Yang, and
  Rebentrost}}]{zhao2021quantum}
\bibinfo{author}{\bibfnamefont{L.}~\bibnamefont{Zhao}},
  \bibinfo{author}{\bibfnamefont{S.}~\bibnamefont{Yang}}, \bibnamefont{and}
  \bibinfo{author}{\bibfnamefont{P.}~\bibnamefont{Rebentrost}},
  \bibinfo{journal}{arXiv preprint arXiv:2109.01014}  (\bibinfo{year}{2021}).

\bibitem[{\citenamefont{Nielsen and Chuang}(2010)}]{nielsen2010quantum}
\bibinfo{author}{\bibfnamefont{M.~A.} \bibnamefont{Nielsen}} \bibnamefont{and}
  \bibinfo{author}{\bibfnamefont{I.~L.} \bibnamefont{Chuang}},
  \emph{\bibinfo{title}{Quantum computation and quantum information}}
  (\bibinfo{publisher}{Cambridge university press}, \bibinfo{year}{2010}).

\bibitem[{\citenamefont{Bresler}(2015)}]{bresler2015efficiently}
\bibinfo{author}{\bibfnamefont{G.}~\bibnamefont{Bresler}}, in
  \emph{\bibinfo{booktitle}{Proceedings of the forty-seventh annual ACM
  symposium on Theory of computing}} (\bibinfo{year}{2015}), pp.
  \bibinfo{pages}{771--782}.

\bibitem[{\citenamefont{Klivans and Meka}(2017)}]{klivans2017learning}
\bibinfo{author}{\bibfnamefont{A.}~\bibnamefont{Klivans}} \bibnamefont{and}
  \bibinfo{author}{\bibfnamefont{R.}~\bibnamefont{Meka}}, in
  \emph{\bibinfo{booktitle}{2017 IEEE 58th Annual Symposium on Foundations of
  Computer Science (FOCS)}} (\bibinfo{organization}{IEEE},
  \bibinfo{year}{2017}), pp. \bibinfo{pages}{343--354}.

\bibitem[{\citenamefont{Kakade et~al.}(2011)\citenamefont{Kakade, Kanade,
  Shamir, and Kalai}}]{NIPS2011_30bb3825}
\bibinfo{author}{\bibfnamefont{S.~M.} \bibnamefont{Kakade}},
  \bibinfo{author}{\bibfnamefont{V.}~\bibnamefont{Kanade}},
  \bibinfo{author}{\bibfnamefont{O.}~\bibnamefont{Shamir}}, \bibnamefont{and}
  \bibinfo{author}{\bibfnamefont{A.}~\bibnamefont{Kalai}}, in
  \emph{\bibinfo{booktitle}{Advances in Neural Information Processing
  Systems}}, edited by
  \bibinfo{editor}{\bibfnamefont{J.}~\bibnamefont{Shawe-Taylor}},
  \bibinfo{editor}{\bibfnamefont{R.}~\bibnamefont{Zemel}},
  \bibinfo{editor}{\bibfnamefont{P.}~\bibnamefont{Bartlett}},
  \bibinfo{editor}{\bibfnamefont{F.}~\bibnamefont{Pereira}}, \bibnamefont{and}
  \bibinfo{editor}{\bibfnamefont{K.}~\bibnamefont{Weinberger}}
  (\bibinfo{publisher}{Curran Associates, Inc.}, \bibinfo{year}{2011}),
  vol.~\bibinfo{volume}{24}.

\end{thebibliography}

\appendix

\section{Proof for ferromagnetic RBM}\label{appendix_A}

We provide detailed proof of Lemma \ref{lemma_rbm_prob} here.

\begin{proof}
We begin by illustrating the procedure to prepare
the state in Eq.~(\ref{equ_probability_preparation}).  By Lemma \ref{lemma_FRBM_index_X_S_equal_to_1}, we have set a $M_S$ that contains indexes of samples where the configuration of each visible node in $S$ is $1$. We prepare the quantum state in Eq.~(\ref{equ_probability_preparation}) using the following  quantum procedure: 
\begin{enumerate}[1).]
\item For each $j\in [n]\setminus (S\cup \{u\})$, and each sample $i\in [M_S]$,  query the quantum access to the samples according to Definition \ref{defQuantumSamples}. This allows us to prepare the state as follows: 
\begin{eqnarray*}
 \ket j \ket {\bar 0}  \to   \ket j \otimes_{i\in[M_S]} \ket{x^{i}_j }, 
\end{eqnarray*}
 where $\ket{x^{i}_j}$  stores the configuration of visible node $j$ in the $i$-th sample in $M_S$.
\item Add a register in state $\ket{\bar{0}},$ apply $M_S$ control-sum gates and store the sum of all state $x_j^i=1$ in the last register. Then, by dividing the sum by $M$, we have
\begin{eqnarray*}
 & & \ket j \otimes_{i\in [M_S]} \ket{x^{i}_{ j } } \ket{\bar{0}}\nonumber\\
 & \to  &  \ket j \otimes_{i\in [M_S]} \ket{x^{i}_{ j } } \ket{\widehat{P}(X_{S \cup \{j\}}=\{1\}^{s+1} ) },
\end{eqnarray*}
where $\widehat{P}(X_{S \cup \{j\}}=\{1\}^{s+1} ) = \frac{1}{M}\sum_{i\in [M_S]} \mathbb I_{\{x^{i}_{ j }=1\}}.$
\item  Undo step 1) for all $i\in[M_S]$, resulting in the state in Eq.~(\ref{equ_probability_preparation}).
\end{enumerate}

Regarding the time complexity, Step $1)$ in the quantum procedure requires queries $\Ord{M_S}$ according to Definition \ref{defQuantumSamples}.  Step $2)$ takes time $\Ord{M_S}$ since it requires $M_S$ controlled-sum gates to calculate the sum and $\Ord{1}$ time to perform the division. Step $3)$ requires  $\Ord{ M_S}$ queries as it is the inverse operation of step 1).  Therefore, the total number of quantum queries is $\Ord{M} $ and the run time is $\Ord{M}$ as $M_S\leq M$.
\end{proof}

\section{Details}

\subsection{Details of obtaining Eq.~(\ref{main_eqcov})}\label{appendix_cov}

Expand the empirical average conditional covariance in Eq.~(\ref{empi_ave_cov}) as 
\begin{eqnarray}
  \widehat{\text{Cov}}^{\text{avg}}(u,v \vert S) &=& \widehat{\mathbb{E}}_{x_S}[ \widehat{\mathbb{E}}[X_uX_v\vert x_S]]
  \nonumber\\
 & -&\widehat{\mathbb{E}}_{x_S}[\widehat{\mathbb{E}}[X_u\vert x_S]\widehat{\mathbb{E}}[X_v\vert x_S]].\label{eq_cov_average}
\end{eqnarray}
We first consider the two terms separately. Let
\begin{eqnarray}
  A_1 & := & \widehat{\mathbb{E}}_{x_S}[ \widehat{\mathbb{E}}[X_uX_v\vert x_S]] \nonumber\\
  A_2 & := & \widehat{\mathbb{E}}_{x_S}[\widehat{\mathbb{E}}[X_u\vert x_S]\widehat{\mathbb{E}}[X_v\vert x_S]],
\end{eqnarray} 
 For the sake of simplicity, we use $P(x_u)$ as the shorthand of  $P(X_u=x_u).$ For a subset $S\in [n]$, as defined in Eq.~(\ref{definition_L_S_F}), $L_S$ is the unique configuration set in the $H$ samples, and set $F(x_S^l)$ is  the sample index set of configuration $x_S^l$ for $ {l\in [\vert L_S \vert] }$.  
We have 
\begin{eqnarray}
  A_1 & =&  \widehat{\mathbb{E}}[ X_u,X_v]
   =  \frac{1}{H} \sum_{i\in[H]} x_u^{i}x_v^{i}=\frac{1}{H} \sum_{i\in[H]} z_{u,v}^{i} \label{term_1},~~~~
\end{eqnarray}
where $x_S\in L_s$, and the first equality is obtained by using the law of total expectation. The third line is obtained by using  Bayes’s theorem and the definition of the marginal distribution.

For the second term $A_2$, we have
\begin{eqnarray}
 A_2 &=& \sum_{x_u,x_v}x_ux_v \mathbb E_{x_S}[\widehat{P}(x_u\vert x_S)\widehat{P}(x_v\vert x_S)]\nonumber\\
&=& \sum_{x_u,x_v,x_S}x_ux_v \widehat{P}(x_u\vert x_S)\widehat{P}(x_v\vert x_S)\widehat{P}(x_S)\nonumber\\
&=& \sum_{x_u,x_v,x_S}\frac{x_ux_v\widehat{P}(x_u,x_S)\widehat{P}(x_v, x_S)}{\widehat{P}(x_S)}, \label{equ2}
\end{eqnarray}
where the third line is obtained by using  Bayes’s theorem and the definition of the marginal distribution.
For Eq.(\ref{equ2}),  we consider $x_u$ and $x_v$ separately. For $j=u,v$, we have 
\begin{eqnarray*}
& & \sum_{x_j} x_j \widehat{P}(x_j, x_S)\nonumber\\
&=&  \widehat{P}(X_j=1, x_S)- \widehat{P}(X_j=-1, x_S)\nonumber\\
&=& \frac{1}{H}  \sum_{i=1}^H  \mathbbm 1_{\{x^{i}_j=1 ,x^{i}_S=x_S\} }-\frac{1}{H}  \sum_{i=1}^H  \mathbbm 1_{\{x^{i}_j=-1 ,x^{i}_S=x_S\} } \nonumber\\
&=&\frac{1}{H}   \sum_{i=1}^H x^{i}_j \mathbbm 1_{\{x^{i}_S=x_S\} }.
\end{eqnarray*}
For a subset $S\in [n]$, as defined in Eq.~(\ref{definition_L_S_F}), $L_S$ is the unique configuration set in the $H$ samples, and $F(x_S^l)$ is the sample index set for configuration $x_S^l$.  
For each $x_S^{l}$, we can obtain
\begin{eqnarray}
a_{j,x_S^{l}} =\sum_{i\in F(x_S^{l})} x^{i}_j,  ~~\text{for}~~j=u,v.
\end{eqnarray}
The term $A_2$ can then be obtained as follows
\begin{eqnarray}
A_2&= & \sum_{x_u,x_v,x_S} \frac{x_u \widehat{P}(x_u, x_S) x_v\widehat{P}(x_v, x_S)}{\widehat{P}( x_S)} \nonumber\\
&=& \frac{1}{H} \sum_{l=1}^{\vert L_S\vert}  \frac{ a_{u,x_S^{l}} a_{v,x_S^{l} }}{ \left\vert F(x_S^l) \right\vert}.\label{term_2}
\end{eqnarray}
Combine Eq.~(\ref{term_1}) and Eq.~(\ref{term_2}) together, the empirical average conditional covariance in Eq.~(\ref{eq_cov_average}) can be written as Eq.~(\ref{main_eqcov}).

\subsection{Proof of Lemma \ref{lemma_quantum_covariance} }\label{appendix_proof_lemma_cov}
We  provide detailed proof of Lemma \ref{lemma_quantum_covariance} in the following.

\begin{proof}

We will now explain the procedure for preparing the state in Eq.~(\ref{equlemma1}). Our strategy is to prepare the two terms in the parentheses separately and then combine them together. The procedure is as follows:

\begin{itemize}
\item[1).] We first consider preparing for the first term. Using the quantum query access, we can prepare the state as follows:
\begin{eqnarray}
\ket{u}\ket{v} \ket{\bar{0}}\rightarrow  \ket{u}\ket{v}\ket{x_u}\ket{x_v},
\end{eqnarray}
where $\ket{x_u} = \otimes_{i=1}^{H}\ket{x_u^{i}},$ $\ket{x_v} = \otimes_{i=1}^{H}\ket{x_v^{i}}$ represents the configurations of $u$ and $v$ for the $i$-th sample, respectively.

We compute the first term in parentheses of Eq.~(\ref{main_eqcov}) via $H$ gates on the first three registers which multiply $x_u$ and $x_v$ and store the result in the third register, followed by  $\Ord{H}$ sum gates. We store the sum of the $H$ products in the last register 
\begin{eqnarray}
& & \ket{x_v} \ket{x_u}\ket{\bar{0}} \ket{\bar{0}} \nonumber\\
&\rightarrow &   \ket{x_v} \ket{x_u}
\left(\otimes_{i=1}^{H}\ket{z_{u,v}^{i} } \right)  \ket{\bar{0}}\nonumber\\
&\rightarrow &  \ket{x_v} \ket{x_u}
\left(\otimes_{i=1}^{H}\ket{z_{u,v}^{i} } \right)  \ket{ \textstyle\sum_{i=1}^{H} z_{u,v}^{i}},\label{eq_term_1_preparation}
\end{eqnarray}
where $\ket{z_{u,v}^{i} }$ stores the product of configurations of $u$ and $v$ for the $i$-th sample, $z_{u,v}^{i} =  x_{u}^{i}x_{v}^{i}$.
\item[2).] Let us now turn to the second term in the parentheses in  Eq.~(\ref{main_eqcov}).  Based on the assumption that the unique configuration set $L_S$ and the sample index subset of each configuration in this set are known (as stated in Lemma \ref{unique_S_via_quantu_access}),  we can prepare $a_{j,x_S^{l}}$ in a quantum state for $j=u,v$ and each $l\in[\vert L_S \vert]$ using $\Ord{H}$  control-sum gates,  according to Eq.~(\ref{eq_a_j_X_S}),
\begin{eqnarray}
\ket{x_j} 
\ket{\bar{0}} 
 \rightarrow  \ket{x_j} 
\textstyle \otimes_{l=1}^{\vert L_S\vert}\ket{ a_{j,x_S^{l}}}, 
\end{eqnarray}
 For example, if the simple index set of configuration $x_S^{1}$ is $F(x_S^{1})=\{1,3,7\}$,  then $\ket{a_{v,x_S^{1}} }=\ket{x^{1}_v+x^{3}_v+x^{7}_v}$ can be obtained using controlled sum gates. By performing a simple calculation, we can prepare the state
\begin{eqnarray}
 \ket{x_v} \ket{x_u}
  \ket {  \sum_{l=1}^{\vert L_S\vert} \frac{a_{v,x_S^{l}}  a_{v,x_S^{l}}  }{ \left\vert {F}\left(x_S^{l}\right)\right\vert  } },\label{eq_term_2_preparation}
\end{eqnarray}
 in time $\Ord{\vert L_S \vert}$. 

\item[3).]Next, we combine the two terms in Eq.~(\ref{eq_term_1_preparation}) and Eq.~(\ref{eq_term_2_preparation}) together according to Eq.~(\ref{main_eqcov}). Then we reverse steps 2) and 1), which leads us to the result in Eq.~(\ref{equlemma1}). 
\end{itemize}

We now analyze the run time. Step $1)$ requires $\Ord{{H}}$ quantum queries  by Definition \ref{defQuantumSamples}. It costs $\Ord{H}$ run time since it involves $\Ord{H}$  gates.   In step $2)$, obtaining $\{ a_{j,x_S^l}\}_{l\in[\vert L_S\vert]}$  for $j=u,v$  requires  $\Ord{H}$ gates, and the remaining calculations take  $\mathcal{O}(\vert L_S\vert)$   where $\vert L_S\vert \in \mathcal{O}(H)$ since $\vert L_S\vert \leq H$.  Therefore, Step 2) requires $\mathcal{O}(H)$ quantum queries, and the total runtime is $\mathcal{O}(H)$.
\end{proof}

\end{document}